\documentclass[11pt]{article}
\usepackage{a4,color,graphics,amsmath,amsfonts,amstext,amssymb,fancyhdr,epsfig}
\usepackage[margin=3cm]{geometry}
\usepackage[english]{babel}
\usepackage{bm}

\begin{document}

\markboth{D. de Falco , D. Tamascelli}
{Quantum Timing and Synchronization Problems}

\title{Quantum Timing and Synchronization Problems
}

\author{
\small{Diego de Falco}\\
\small{Dipartimento di Scienze dell'Informazione}\\
\small{Universit\`a degli Studi di Milano}\\
\small{via Comelico 39, 20135 Milano, Italy}\\
\small{\itshape{e-mail: defalco@dsi.unimi.it}}
\and
\small{Dario Tamascelli}\\
\small{Dipartimento di Matematica}\\
\small{Universit\`a degli Studi di Milano}\\
\small{via Saldini 50, 20133 Milano, Italy}\\
\small{\itshape{e-mail: tama@mat.unimi.it}}}
\date{}
\maketitle

\begin{abstract}
Feynman's model of a quantum computer provides an example of a continuous-time quantum walk. Its clocking mechanism is an excitation of a basically linear chain of spins with occasional controlled jumps which allow for motion on a planar graph. The spreading of the wave packet poses limitations on the probability of ever completing the $s$ elementary steps of a computation: an additional amount of storage space $\delta$ is needed in order to achieve an assigned completion probability. In this note we study the END instruction, viewed as a measurement of the position of the clocking excitation: a $\pi$-pulse indefinitely freezes the contents of the input/output register, with a probability depending only on the ratio $\delta/s$.
\end{abstract}


\section{Introduction} \label{S:intro}

It has been shown by Feynman\cite{Feynman 86} that it is possible to implement the sequential application, in the desired order, of a collection $ A = A_{s-1} \cdot \dots \cdot A_2 \cdot A_1$ of unitary operators to an \emph{input/output register} by using \emph{s} additional \emph{program counter sites}.\\
For the sake of definiteness, we will think of each program counter  site $j=1,2, \dots, s$ as occupied by a spin-1/2 system $\underline{\tau}(j)=(\tau_1(j),\tau_2(j),\tau_3(j))$. We will refer to the collection of such spins, which act in effect as a quantum clocking mechanism, as to a ``\emph{program line}''.\\ 
The \emph{input/output register} will be, similarly, implemented by a collection of a certain number $\mu$ \ of spin-1/2 systems $\underline{\sigma}(i)=(\sigma_1(i),\sigma_2(i),\sigma_3(i))$, $ i=1,2, \dots,\mu $.\\       
The evolution of the \emph{complete} system, \emph{register + program line}, will be given by the Schr\"odinger equation:

\begin{equation}
i \frac{d}{dt} | \psi(t) \rangle = H | \psi(t) \rangle,
\label{eq 1}
\end{equation}

where the Hamiltonian $H$ is supposed to be time independent and involving at most 3-body interactions:

\begin{equation}
H=-\frac{1}{2}\Bigl ( \sum_{j=1}^{s-1} \tau_+(j+1)A_j \tau_-(j)+ hermitian \; conjugate\Bigr ).
\label{E:firsth}
\end{equation}

We have indicated by $\tau_+(j)$ and $\tau_-(j)$ the raising and lowering operators for the third component of the spin occupying the $j$-th program counter site.\\
Since it is requested that each term of the sum is (at most) a 3-body interaction, each $A_j$  either acts on a single spin of the register or is a constant: as shown in Ref.\cite{Feynman 86}, this does not restrict the class of functions computable by the model.

We will restrict our considerations to initial conditions of the form:

\begin{equation}
|\psi(0)\rangle = | register_0 \rangle \otimes | program\; line_0 \rangle,
\label{E:init}
\end{equation}

where $| program\; line_0 \rangle=| \tau_3(1)=+1, \tau_3(j)=-1 \mbox{ for } 1< j \leq s \rangle$ describes the state in which \underline{only} the spin located at program site number 1 is ``up''.\\
It helps the intuition to think of the initial state $| register_0 \rangle$  of the register as a simultaneous eigenstate of the components of the $\sigma$ spins in selected directions, encoding the initial word (or superposition of words) on which the machine is required to act.

Because of the conservation law $[H,N_3]=0$, where 

\begin{equation}
N_3=\sum_{j=1}^{s} \frac{1+\tau_3(j)}{2},
\end{equation}

the above initial condition has the important consequence that the evolution $| \psi(t) \rangle$ takes place in the  $2^\mu s$ dimensional eigenspace of $N_3 $  belonging to the eigenvalue +1.\\
The intuition of ``a single clocking excitation travelling along the program line'' emerging from the above conservation law is made precise by introducing the observable \emph{position of the excitation}, or \emph{position of the cursor}:

\begin{equation}
Q=\sum_{j=1}^{s}j \; \frac{1+\tau_3(j)}{2}.
\end{equation}

It is then easy to convince oneself that the evolution of the overall system is of the form 

\begin{equation}
| \psi(t) \rangle= \sum_{k=1}^s c(t,k;s) \; | register_{k-1} \rangle \otimes | Q=k \rangle
\label{E:evolution}
\end{equation}
   
where, for $1 \leq h \leq s-1$, $|register_h \rangle = A_h | register_{h-1} \rangle$.\\
In Feynman's words (adapted to our notations), (\ref{E:evolution}) says that, starting from the initial condition (\ref{E:init}), ``\underline{If} at some later time the final site $s$ is found to be in the  $| \tau_3(s)=+1 \rangle$  state (and therefore all the others in $| \tau_3(j)=-1 \rangle$ ), then the register state has been multiplied by $A_{s-1}\cdot \dots \cdot A_2 \cdot A_1$  as desired''. It as been shown in Ref.\cite{Apolloni 01} that this is a somewhat big ``If'', under two respects:

\begin{enumerate}
\def\theenumi{\roman{enumi}}
\item at no instant of time the probability $| c(t,s;s) |^2$ is larger than $const \cdot s^{-\frac{2}{3}}$ ; \label{R:i}
\item the cursor keeps bouncing back and forth between positions 1 and $s$, thus in effect making the above upper bound attainable only at selected instants of time.\label{R:ii}
\end{enumerate}

\noindent The above two statements are reviewed and made quantitative in Section \ref{S:cursor}.\\
Section \ref{S:pipulse} is devoted to the ``quantum END problem'': we remove the cursor in order to prevent it from returning down the program line and ``undoing'' the computation. Removing the cursor and storing the result of the computation in the contents of the register is in effect a measurement procedure, that in Sec. \ref{S:pipulse} will be modelled by a suitable time dependent perturbation (a $\pi$-pulse) applied to a variant of the Hamiltonian (\ref{E:firsth}).\\
Section \ref{S:conclusions} is devoted to conclusions and outlook.

\section{The Motion of the Cursor} \label{S:cursor}

We recall, first of all, that the motion of the cursor does not depend on the operators  acting on the register.\cite{Feynman 86} For the particular case of a {\it sequential program line} as the one described by the Hamilonian (\ref{E:firsth}), this is made evident by the explicit expression\cite{Gramss 95} of the amplitudes $c(t,k;s)$ in (\ref{E:evolution}): they are given, independently of the operators $A_j$, by:

\begin{equation}
c(t,k;s)=\frac{2}{s+1} \sum_{n=1}^s \exp {(i t \cos{(\theta(n;s))})}\sin{(\theta(n;s))}\sin{(k \, \theta(n;s))},
\label{gramsscoeff}
\end{equation}

where:

\begin{equation}
\theta(n;s)=\frac{n\pi}{s+1}
\end{equation}

Similar results hold in the case, studied in Ref.\cite{de Falco 03}, in which, because of {\it conditional jumps} in the program line (such as the ones needed in the iteration of {\it quantum subroutines}), the cursor performs, in effect, a continuous-time quantum walk\cite{Childs 02} on a planar graph. In this note we restrict ourselves to the {\it sequential} case.\\
The main purpose of this Section is to give examples of the behaviour recalled in the observations (\ref{R:i}) and (\ref{R:ii}) of Sec.\ref{S:intro}.\\
This we do with the help  of the following Hamiltonian:
\begin{eqnarray}
\label{E:htelchain}
H& = & -\frac{1}{2} \Bigl( \sum_{j=1}^{s-1} \tau_+(j+1)A_j \tau_-(j)+ \nonumber \\
 &   & +\tau_+(s+1) \rho_- \tau_-(s)+ \nonumber \\
 &   & +\sum_{j=s+1}^{s+\delta-1}\tau_+(j+1)\tau_-(j)+ h.c. \Bigr ).
\end{eqnarray}

\begin{figure}[h]
\vspace{1cm}
\hspace{3cm}
\begin{picture}(180,130)(0,0)
\centering
\includegraphics[width=8cm]{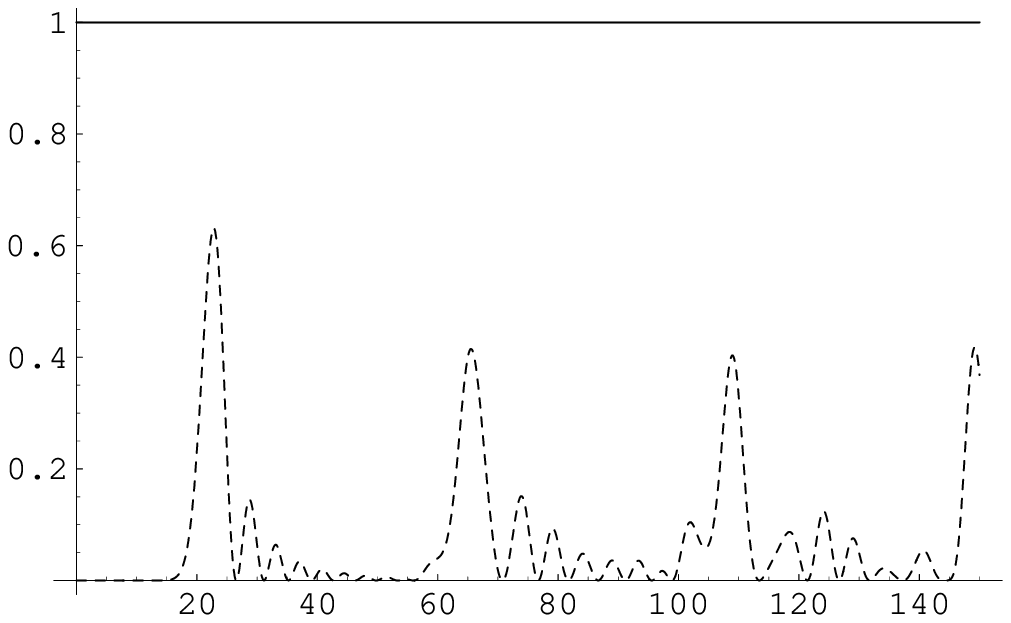}
\end{picture}
\begin{picture}(10,130)
\put(50,11){t}
\end{picture}
\caption{$|program \; line_0 \rangle = |\rho_3=-1 \rangle \otimes | Q=1 \rangle$; $s=20$.}
\label{F:cursor}
\end{figure}

\begin{figure}[!h]
\hspace{3cm}
\begin{picture}(180,130)(0,0)
\centering
\includegraphics[width=8cm]{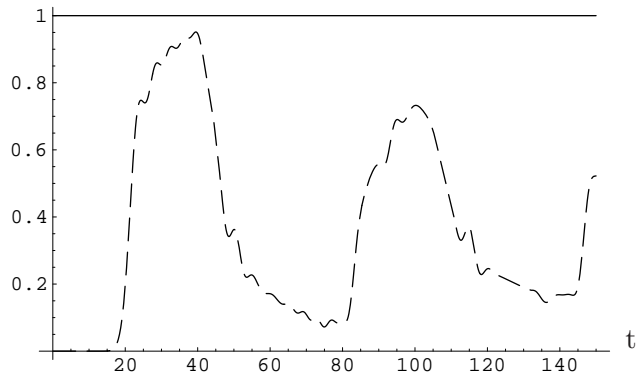}
\end{picture}
\begin{picture}(10,130)
\put(50,11){t}
\end{picture}
\caption{$|program\; line_0\rangle= |\rho_3=+1\rangle \otimes |Q=1 \rangle$; $s=20$; $\delta =10$.}
\label{F:telo}
\end{figure}

With respect to the Hamiltonian (\ref{E:firsth}), we have introduced an additional {\it control} q-bit $\underline{\rho}=(\rho_1,\rho_2,\rho_3)$ in the term $\tau_+(s+1)\rho_- \tau_-(s)$; this is an example of a \emph{conditional jump} in the quantum walk performed by the cursor: it acts non trivially only in the eigenspace belonging to the eigenvalue +1 of $\rho_3$, \emph{enabling}  the transition  $| Q=s \rangle \rightarrow | Q = s+1 \rangle $. If this transition is enabled, then the cursor can visit the additional {\it telomeric} sites $s+1,\dots, s+\delta$, else it gets reflected back.\\
Figures \ref{F:cursor} and \ref{F:telo} give examples of the behaviour of the probability

\begin{equation}
P_{(s \leq Q)}(t)=P_{(s\leq Q \leq s+\delta)}(t)=\sum_{j=s}^{s+\delta} | \gamma(t,j) |^2
\label{E:Probchain}
\end{equation}

of finding the register in the state $A = A_{s-1}\cdot \dots \cdot A_2 \cdot A_1 |register_0 \rangle$, under two different initial conditions, which determine two different forms of the amplitudes~$\gamma$.\\
Figure \ref{F:cursor} corresponds to the initial condition \mbox{$|program\;line_0\rangle$} = \mbox{$|\rho_3=-1\rangle \otimes$}\mbox{$| Q=1 \rangle $}: the motion of the cursor remains confined to the sites $1,\dots,s$, as it is $\gamma(t,k)=c(t,k;s)$ if $1 \leq k \leq s$, 0 otherwise. The probability $P_{(s \leq Q)}(t)$ of finding the computation completed satisfies in this case the inequality:\cite{Apolloni 01}
\begin{equation}
P_{(s \leq Q)}(t) \leq \frac{8.}{s^{\frac{2}{3}}}
\label{E:probnotelo}
\end{equation}
Figure \ref{F:telo} corresponds to the initial condition: $|program\;line_0\rangle =|\rho_3=+1\rangle \otimes$ \mbox{$| Q=1 \rangle$}, leading to $\gamma(t,k)=c(t,k;s+\delta)$ for $1 \leq k \leq s+\delta$. For $t$ just below $s+2\delta$ the probability $P_{(s \leq Q)}(t)$ of finding the computation completed is close to the much less severe upper bound:\cite{Apolloni 01}
\begin{equation}
P_{(s \leq Q \leq s+\delta)}(t) \leq 1-\frac{2}{\pi} \Biggl (  \arcsin \Bigl (  \frac{1}{1+2 \delta/s}\Bigr )- \Bigr (   \frac{1}{1+2\delta/s} \Bigr ) \sqrt{1- \Bigl (\frac{1}{1+2 \delta/s} \Bigr ) ^2 }\Biggr).
\label{E:upperbound}
\end{equation}

\section{The Quantum END Instruction} \label{S:pipulse}
\begin{figure}[!hb]
\vspace{1cm}
\hspace{2.7cm}
\begin{picture}(180,130)(0,0)
\centering
\includegraphics[width=9cm]{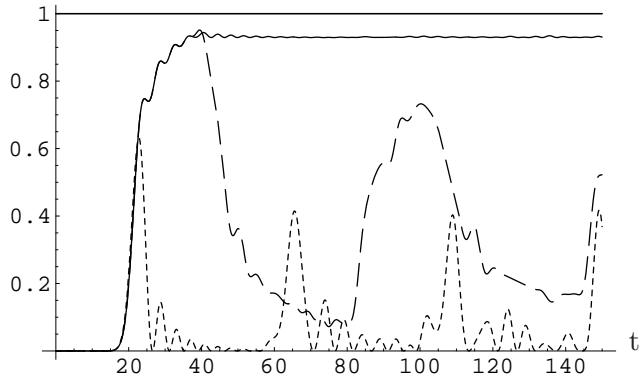}
\end{picture}
\begin{picture}(10,130)
\put(60,11){t}
\end{picture}
\caption{ The solid line represents the probability of finding the cursor in the telomeric chain using a $\pi$-pulse applied at time $t_0=s+2\delta$. The dashed lines correspond to Figs.\ref{F:cursor} and \ref{F:telo}.}
\label{F:pipulse}
\end{figure}

The abrupt collapse of $P_{(s \leq Q)}(t)$ at time $t\approx s+2\delta$, evident from Fig. \ref{F:telo}, corresponds to the following fact: travelling with average speed close to 1, at time $t \approx s+ 2\delta$ the cursor ``returns down the active part of the program line'', thus, in effect, undoing the calculation.\\
Bringing the computation to an END, and storing the result is not completely trivial in the case examined here of a reversible quantum clocking mechanism: ``Surely a computer has eventually to be \underline{in interaction with the external world}, both for putting data in and for taking it out\cite{Feynman 86}''.\\
A simple model of such interaction is suggested by inspection of the Hamiltonian (\ref{E:htelchain}): starting from the initial condition $| \rho_3=+1 \rangle$, the transition \mbox{$| Q=s \rangle$} \mbox{$\rightarrow$} \mbox{$|Q=s+1 \rangle$} is enabled by the control term $\tau_+(s+1)\rho_- \tau_-(s)$ which, simultaneously, determines the transition $|\rho_3=+1 \rangle \rightarrow |\rho_3=-1 \rangle$.\\
The transition $|Q=s+1 \rangle \rightarrow |Q=s \rangle$, enabled by the hermitian conjugate term $\tau_+(s)\rho_+ \tau_-(s+1)$, will be therefore inhibited if, by external means, we enforce the transition $|\rho_3=-1 \rangle \rightarrow |\rho_3=+1 \rangle$ at a time, close to $t_0=s+2\delta$, when most of the probability mass is in the region $s,\dots, s+\delta$.\\
Figures \ref{F:pipulse} (where Figs. \ref{F:cursor} and \ref{F:telo} are also reproduced for comparison purpose) presents the effect of the addition to (\ref{E:htelchain}) of the time dependent perturbation
\begin{equation}
h(t)=\frac{1}{2}B(t)\cdot \rho_1
\label{E:pipulse}
\end{equation}
where the ``magnetic field'' $B(t)$ is non vanishing only in a unit time interval around $t_0$, in which it takes the value $\pi$.
With a probability depending only on the ratio $\delta/s$ (see (\ref{E:upperbound})) between the lengths of the active part and the telomeric part of the program line, the $\pi$-pulse (\ref{E:pipulse}) definitively prevents the cursor from undoing the computation.\\
The idea of a $\pi$-{\it pulse trap} just presented works only if the control q-bit is initialised in the $|\rho_3=+1\rangle$ state. It is immediate to convince oneself that the following {\it double trap} Hamiltonian does not suffer from the above limitation:
\begin{eqnarray}
\label{E:htelfork}
H& = &h(t)+ \nonumber \\
 &   &-\frac{1}{2} \Bigl( \sum_{j=1}^{s-1} \tau_+(j+1)A_j \tau_-(j)+\sum_{j=s+1}^{s+\delta-1}\tau_+(j+1)\tau_-(j)+\sum_{j=s+\delta+1}^{s+2\delta-1}\tau_+(j+1)\tau_-(j)+
  \nonumber \\
 &   &+\tau_+(s+1) \rho_- \tau_-(s)+ \tau_+(s+\delta+1) \rho_+ \tau_-(s) +h.c. \Bigr ).
\end{eqnarray}

With any initial condition for the control q-bit, under the action of the above Hamiltonian, the $|\rho_3=+1\rangle$ component of the state gets definitively trapped in the first telomeric region $\{s+1,\dots,s+\delta \}$, the $|\rho_3=-1 \rangle$ component in the second one $\{s+\delta+1,\dots,s+2\delta \}$.\\[4pt]
As a final remark of this section, we observe that, acting in effect as a Stern-Gerlach apparatus providing \emph{space} separation between two different \emph{spin} states, the term
\begin{equation}
switch=(\tau_+(s+1) \rho_- \tau_-(s)+ \tau_+(s+\delta+1) \rho_+ \tau_-(s))+h.c.
\label{E:switch}
\end{equation}

can be used also to model the preparation (``putting the data in'') of a {\it register} q-bit in a given spin state.

\section{Conclusions and Outlook}\label{S:conclusions}
Feynman's time honoured model of a quantum computer and its modern streamlined version, the continuous-time quantum walk\cite{Childs 02} paradigm (in which the quantum motion of the {\it cursor} on a graph \underline{is} the computation, irrespective of any action on the register), provide a fascinating physical context in which to think of ``time'' under a quantum perspective and are rich sources of open problems.\\
Inequality (\ref{E:probnotelo}) is, for istance, a simple consequence of the spreading of the wave packet in the inertial motion (\ref{gramsscoeff}) of the cursor on a finite lattice. Is inequality (\ref{E:probnotelo}) strictly dependent on the model adopted here or is it a hint of a model independent ``probability vs. computational complexity'' uncertainty relation?\\
Inequality (\ref{E:upperbound}) and the related discussion of Sec.~\ref{S:pipulse} set bounds on the minimun amount of additional space needed in order to have a preassigned probability of storing the result of the complete calculation. Similarly, the preparation of a given input requires time and space in order to perform the required ``Stern-Gerlach'' preparation. Is it realistic to neglect this costs as it is done in the conventional performance analysis of performance of quantum algorithms?\\[4pt]
As a concluding remark, we wish to point out a case, Grover's algorithm\cite{Grover 96}, in which the nature, classical or quantum, of the clocking agent that successively applies the required primitives does make a difference.\\
Grover's algorithm poses the problem of estimating the parameters $\mathbf{a}=(a_1,\dots,a_{\mu} )\in \{-1,+1\}^{\mu}$ appearing in the transformation $\mathbf{A}=1-2 P_\mathbf{a}$ that an \emph{oracle} is able to apply to the register. Here we have indicated by $P_\mathbf{a}$ the projector on the state $|\sigma_3(i)=a_i,\; i=1,\dots,\mu \rangle \equiv | \mathbf{a}_3 \rangle$.\\
The estimation procedure of {\bf a} starts from $|register_0 \rangle=$ \mbox{$| \sigma_1(i)=+1,\; i=1,\dots,\mu \rangle$} $\equiv | \mathbf{1}_1 \rangle$ and proceeds by alternating the action of {\bf A} with the action of $\mathbf{B}=1-2 |\mathbf{1}_1 \rangle \langle \mathbf{1}_1 |$.\\
It can be shown\cite{Boyer 98} that:

\begin{equation}
\label{E:amplsol}
| \langle \mathbf{a}_3|(\mathbf{B} \cdot \mathbf{A})^n | \mathbf{1}_1 \rangle |^2 =\sin^2((2n+1) \vartheta),\mbox{ with }\vartheta=\arcsin(2^{-\mu / 2}).
\end{equation}

The sharp maximum of (\ref{E:amplsol}) at $n=n_{optimal} \approx \frac{\pi}{4} 2^{\mu /2}$ keeps reappearing periodically if the computation proceeds indefinitely after $n_{optimal}$ steps.\\
In Refs.\cite{Fahri 98} and \cite{Childs 03}, the oscillatory nature of the overlap (\ref{E:amplsol}) between the target state $|\mathbf{a}_3 \rangle$ and the current state $(\mathbf{B} \cdot \mathbf{A})^n |\mathbf{1}_1 \rangle$ has been nicely explained in terms of the spectral gap of an analogue Hamiltonian of the form $- \gamma |\mathbf{a}_3 \rangle \langle \mathbf{a}_3| - |\mathbf{1}_1 \rangle \langle \mathbf{1}_1| $ acting on the \emph{register} viewed as an isolated system.\\[4pt]
In the context of the model considered in the previous sections, Grover's algorithm corresponds to the execution of a \emph{program line} of the form (\ref{E:firsth}) with $A_j = \mathbf{A}$ for odd $j$, $A_j = \mathbf{B}$ for even $j$.\\
Starting form the initial condition $ | \psi(0)\rangle = | \mathbf{1}_1 \rangle \otimes | Q=1 \rangle $, and taking, for the sake of definiteness, an odd value of $s$, $s=2g+1$, it is easy to check that the overlap probability is, in this case, given by
\begin{eqnarray}
\label{E:damping}
\langle\psi(t) | P_{\mathbf{a}} | \psi(t) \rangle & = & \sum_{n=0}^{g} \sin^2((2n+1) \vartheta)\bigl(|c(t,2n+1;s)|^2+ |c(t,2n+2;s)|^2 \bigr)= \nonumber \\ 
& = & \sum_{x=1}^{s} |c(t,x;s)|^2 \sin^2(\vartheta \; x_{odd}),
\end{eqnarray}

where $x_{odd}$ is the largest odd number not larger than $x$.\\
Direct inspection of (\ref{E:damping}) shows that (\ref{E:amplsol}) gives the {\it \underline{conditional}} probability of finding the register in the target state, \emph{given} that the cursor is in the state $| Q= 2n+1 \rangle $; in loose terms: the oscillatory behaviour (\ref{E:amplsol}) describes the overlap probability as a function of the ``machine time'', namely the position $Q$ of the clocking excitation.\\
\begin{figure}[!ht]
\hspace{3cm}
\begin{picture}(180,130)(0,0)
\centering
\includegraphics[width=8cm]{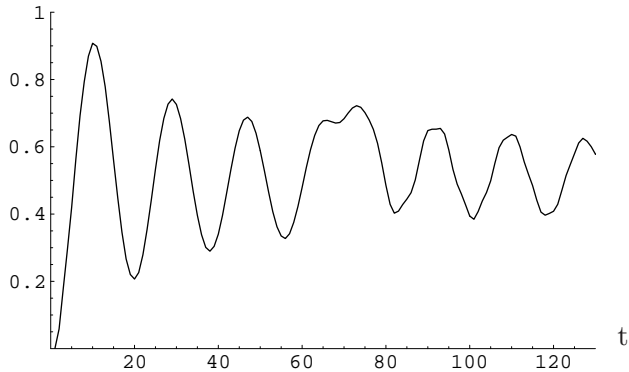}
\end{picture}
\begin{picture}(10,130)
\put(50,11){t}
\end{picture}
\caption{$ \mu=5$; $s=2^{\mu + 1}+1=65$.}
\label{F:damping}
\end{figure}

The right hand side of (\ref{E:damping}) gives, instead, the overlap probability as a function of the ``Galileian time'' $t$, the independent variable in (\ref{eq 1}).\\
Figure \ref{F:damping} shows the behaviour of $\langle\psi(t) | P_{\mathbf{a}} | \psi(t) \rangle $ as a function of $t$: the coupling of the \emph{register} with the \underline{many} clocking degrees of freedom has a damping effect on the oscillation of the state of the register.\\
In the example just discussed we have used, in fact, a number of clocking degrees of freedom growing exponentially with $\mu$. It is shown in Refs.\cite{de Falco 03} and \cite{Tamascelli 03} that the above pseudo-dissipative effect can still be observed if, by a careful use of \emph{quantum subroutines}, the number of program line sites is reduced to a polynomial in $\mu$. 

\section*{Acknowledgements}
It is a pleasure to thank Professor Alberto Bertoni for his constant attention and encouragement during completion of this work.

\end{document}